\documentclass[manuscript]{acmart}
\usepackage{xcolor}
\AtBeginDocument{%
  \providecommand\BibTeX{{%
    \normalfont B\kern-0.5em{\scshape i\kern-0.25em b}\kern-0.8em\TeX}}}

\copyrightyear{2021}
\acmYear{2021}
\setcopyright{rightsretained}
\acmConference[AHs '21]{Augmented Humans International Conference 2021}{February 22--24, 2021}{Rovaniemi, Finland}
\acmBooktitle{Augmented Humans International Conference 2021 (AHs '21), February 22--24, 2021, Rovaniemi, Finland}\acmDOI{10.1145/3458709.3458946}
\acmISBN{978-1-4503-8428-5/21/02}

\usepackage{array,multirow,graphicx}
\begin{document}

\title{Exploratory Design of a Hands-free Video Game Controller for a Quadriplegic Individual}

\author{Atieh Taheri}
\email{atieh@ece.ucsb.edu}
\affiliation{
\institution{University of California,
Santa Barbara}
\city{Santa Barbara, CA 93106}
\country{USA}
}
\author{Ziv Weissman}
\email{zivweissman2008@gmail.com}
\affiliation{
\institution{Palo Alto Senior High School}
\city{Palo Alto, CA 94301}
\country{USA}
}
\author{Misha Sra}
\email{sra@cs.ucsb.edu}
\affiliation{
\institution{University of California,
Santa Barbara}
\city{Santa Barbara, CA 93106}
\country{USA}
}

\renewcommand{\shortauthors}{Taheri et al.}

\newcommand\mcomment[1]{\textcolor{red}{Misha:#1}}

\begin{abstract}
From colored pixels to hyper-realistic 3D landscapes of virtual reality, video games have evolved immensely over the last few decades. However, video game input still requires two-handed dexterous finger manipulations for simultaneous joystick and trigger or mouse and keyboard presses. In this work, we explore the design of a hands-free game control method using realtime facial expression recognition for individuals with neurological and neuromuscular diseases who are unable to use traditional game controllers. Similar to other Assistive Technologies (AT), our facial input technique is also designed and tested in collaboration with a graduate student who has Spinal Muscular Atrophy. Our preliminary evaluation shows the potential of facial expression recognition for augmenting the lives of quadriplegic individuals by enabling them to accomplish things like walking, running, flying or other adventures that may not be so attainable otherwise.
 
\end{abstract}




\begin{teaserfigure}
  \includegraphics[width=\textwidth]{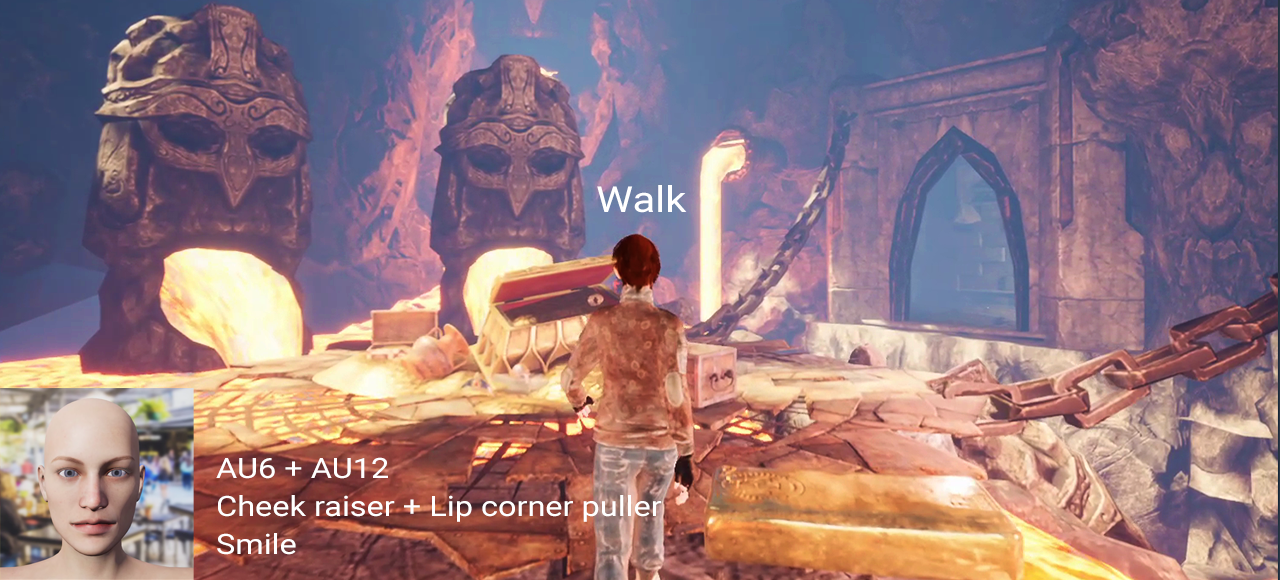}
  \caption{A screenshot from the Temple Looter adventure game. The facial action units corresponding to a smile (AU6 + AU12 or Cheek raiser + Lip corner puller) are recognized from realtime video input data to make the video game character walk.}
  \label{fig:teaser}
\end{teaserfigure}

\maketitle

\section{Introduction}

With COVID-19 pandemic induced stay-at-home lifestyles, video conferencing has become a prime way to connect, whether it is for work, school, fitness or socializing. Interest in video games has also risen as sequestered people around the world take to games not only as an escape from the impositions of the pandemic but also for work related meetings~\footnote{\url{https://www.nytimes.com/2020/07/31/business/video-game-meetings.html}}, and as a new way to connect with family and friends~\footnote{\url{https://www.nytimes.com/2020/04/21/technology/personaltech/coronavirus-video-game-production.html}}.
For a lot of people, video games are about being able to do things that are often not possible in real life, about experiencing great adventures and visiting new places. Yet, as prolific as gaming is, it is inaccessible to a significant number of people with disabilities~\cite{yuan2011game}. According to the CDC, 1 in 4 US adults have some type of disability\footnote{\url{https://www.cdc.gov/media/releases/2018/p0816-disability.html}}).

With growing interest in playing video games and with video games increasingly being used for purposes other than entertainment, such as education~\cite{gee2003video}, rehabilitation~\cite{howcroft2012active,lange2009initial} or health~\cite{warburton2007health,kato2010video}, game accessibility is increasingly critical, and even more so for players with disabilities who stand to benefit greatly from the opportunities video games offer. But, games are usually far more demanding than other entertainment media ``in terms of motor and sensory skills needed for interaction control, due to special-purpose input devices, complicated interaction techniques, and the primary emphasis on visual control and attention''~\cite{grammenos2009designing}.

Individuals with neuromuscular diseases can have difficulties grasping, holding, moving, clicking, pushing or pulling due to lack of muscle control. Loss of muscle function could happen due to degenerative neurological diseases such as Muscular Dystrophy (MD), Spinal Muscular Atrophy (SMA), Multiple Sclerosis (MS), or Amyotrophic Lateral Sclerosis (ALS or Lou Gehrig’s disease), or from brain and spinal cord injuries, for example, due to motor vehicle accidents and stroke. Standard game input devices such as a mouse or a keyboard are not well suited to the needs of users with severe motor disabilities~\cite{cecilio2016bci}. A number of input devices have been developed to allow motor impaired players to interact with games such as mechanical switches~\cite{perkins1986control}, mouth and tongue controllers and joysticks~\cite{krishnamurthy2006tongue,peng2007zigbee,quad}, brain-computer interfaces~\cite{pires2012evaluation}, and eye-gaze controllers~\cite{smith2006use,gips1996eagleeyes}. The type of device that can work depends on an individual's needs and requirements that are determined by the targeted muscles and the degree of function of those muscles. However, most of these devices are typically constrained with regard to the input they can provide when compared with the types of input required in conventional games. 

In this work, we present a hands-free human-computer interaction solution designed in collaboration with a quadriplegic user. The solution is based on realtime facial expression recognition (FER). By recognizing muscle movements in a webcam video stream, facial expressions (FEs) are identified and mapped to a PC keyboard to control actions in a video game. The system also includes speech recognition which serves as a secondary input modality. Unlike brain-computer interfaces or other input technologies designed for motor-impaired users, our system is inexpensive and does not encumber the user with sensors and devices.  
 
The main contribution of this work are: 
\begin{itemize}
    \item A fully functional prototype of facial expression recognition based video game control for individuals with quadriplegia.
    \item Collaborative design approach for design and testing of facial expression recognition.
    \item Design of three different games that demonstrate the mapping of facial expressions to game actions with focus on user agency, user comfort, ease of use, memorability, and reliability of recognition along with some design reflection.
\end{itemize}

\section{Related Work}
There has been a lot of research in the past twenty years aimed at developing assistive technology (AT) devices to increase independence and in individuals with motor impairments of various origins (e.g., locked-in-syndrome, amyotrophic lateral sclerosis, spinal muscular atrophy, quadriplegia, muscular dystrophy, cerebral palsy, etc.)~\cite{pinheiro2011alternative}. We present a subset of that work here that is specifically related to interaction with video games, both for a broader audience of disabled users and for those with  motor disabilities. 

\begin{figure*}[!t]
    \centering
    \includegraphics[width=0.95\textwidth]{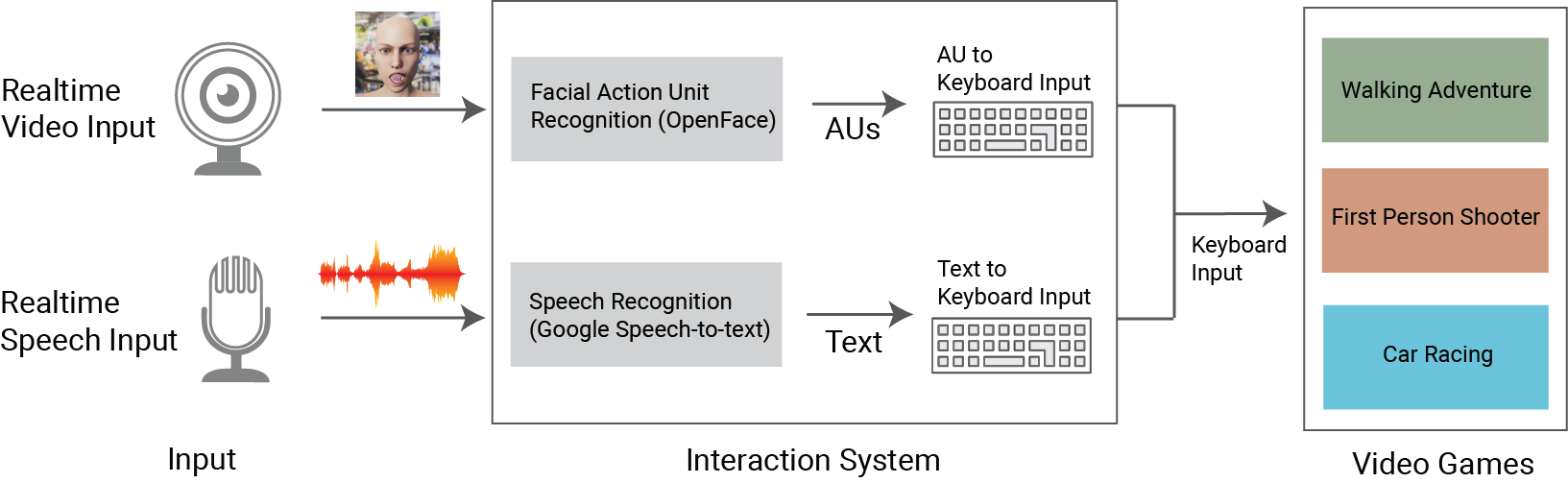}
    \caption{The system pipeline showing input of video and speech data that is processed and converted to keyboard bindings for controlling action in each game.}
    \label{fig:pipeline}
\end{figure*}

\subsection{Accessibility in Consumer Video Games}
More consumer games are starting to include accessibility options. For example, The Last of Us: Part II released in June 2020 offers around 60 accessibility options like directional subtitles and awareness indicators for deaf players or auto-target and auto-pickup for those with motor disabilities~\footnote{\url{https://www.playstation.com/en-us/games/the-last-of-us-part-ii-ps4/accessibility/}}. The accessibility options in Naughty Dog's 2016 release Uncharted 4: A Thief's End~\footnote{\url{https://dagersystem.com/disability-review-uncharted-4/}}, support features like auto-locking the aiming reticle onto enemies, changing colors for colorblind users, or adding help to highlight enemies. Sony has included a number of accessibility functions in the PS4 system\footnote{\url{https://support.playstation.com/s/article/PS4-Accessibility-Settings?language=en_US}}, including text-to-speech, button remapping, and larger font for players with visual and auditory impairments. 

In addition to adding accessibility options in commercial games, many special purpose games have also been developed for blind players~\cite{yuan2008blind,friberg2004audio,morelli2010vi}.
Canetroller~\cite{zhao2018enabling} is device that enables visually impaired individuals to navigate a virtual reality environment with haptic feedback through a programmable braking mechanism and vibrations supported by 3D auditory feedback. Virtual Showdown~\cite{wedoff2019virtual} is a virtual reality game designed for youth with visual impairments that teaches them to play the game using verbal and vibrotactile feedback. Players of a recently release game, Animal Crossing: New Horizons are using the game's customization options to make the game more accessible. For example, a blind player demonstrated how they modified the game in ways that do not rely on sight while another player low-vision player covered their island in grass and flowers to force fossils and rocks to spawn in specific spots~\footnote{\url{https://kotaku.com/how-animal-crossing-new-horizons-players-use-the-game-1844843087}}. Not all commercial games are customizable which leaves some players with disabilities are to rule out those games or seek help of a friend or assistant to ``play'' the game.

The leading example of an accessible game controller is Microsoft's Xbox Adaptive game controller that allows people with physical disabilities who retain hand/finger movement and control, to be able to interact and play games~\cite{xbox}. By connecting the adaptive controller to external buttons, joysticks, switches and mounts, gamers with a broad range of disabilities can customize their setup. The device can be used to play Xbox One and Windows 10 PC games and supports Xbox Wireless Controller features such as button remapping~\cite{Bach01}.

The solutions presented here, while accessible, are not usable by those with severe motor disabilities as most of these solutions rely on hand-based control. Additionally, while both software and hardware solutions can make gaming accessible, we believe software solutions can provide a more economical and customizable solution. Thus, in this work we explore the design of a software based input system.

\subsection{Game Input Methods for Quadriplegics}

For individuals with quadriplegia or those with severe motor impairments, there are a few approaches that can enable playing video games without assistance. Most of these are based on acquiring signal from different parts of the body such as the tongue, brain, or muscles that the individual has voluntary control over. There are several Tongue Machine Interfaces (TMIs) such as  tongue-operated switch arrays~\cite{struijk2006inductive} or permanent magnet tongue piercings that are detected by magnetic field sensors~\cite{krishnamurthy2006tongue,huo2008magneto} to enable interaction with a computer.  Lau et al.~\cite{lau1993comparison} created a radio frequency transmitting device shaped like an orthodontic retainer containing Braille keys that could be activated by raising the tongue tip to the mouth superior palate. Leung et al.~\cite{leung2008multiple} presented a theoretical framework for using a multi-camera system for facial gesture recognition for children with severe spastic quadriplegic cerebral palsy. Chen et. al.~\cite{chen2003computer} mapped eye and lip movements to a computer mouse for a face-based input method. 

Guangyu et. al.~\cite{bin2011high} showed a high-speed spelling system based on a brain-computer interface (BCI). BCIs can provide non-muscular communication and control to people who are severely motor impaired. Non-invasive BCIs using scalp-recorded electroencephalographic (EEG) brain activity along with adaptive algorithms have not become popular among users despite being researched since the early 1970s \cite{vidal1973toward,wolpaw2002brain,birbaumer1999spelling,mcfarland2008emulation} due to limitations like bandwidth and susceptibility to noise and interference~\cite{huo2010evaluation}. 

Buttons or switches are a commonly used device to enable motor impaired players to interact with games as they can be operated with any part of the body that is able to produce voluntary movement, enabling actions like sip and puff, pull, push and squeeze~\cite{yuan2011game}. Depending on the severity of the motor impairment one or more switches may be usable with binary input being the smallest amount of interaction provided with a switch.

There are a few consumer products that allow motor impaired users to play video games. One Switch\cite{oneswitch} is a non-profit dedicated to arcade style games that can be played with one switch. The QuadStick~\cite{quad} is a mouth operated device available in three versions that allow interacting with a computer or playing video games. Depending on the version, it includes sip/puff pressure sensors, a lip position sensor, and a joystick with customizable input and output mapping. Quadstick has enabled players with severe disabilities play videos games and engage in social interaction through streaming on Twitch~\footnote{\url{https://www.washingtonpost.com/video-games/2019/10/14/its-my-escape-how-video-games-help-people-cope-with-disabilities/}}. Axis controllers~\cite{axis} enable players who are able to move their wrists to play console games with an ergonomic layout featuring large buttons and joysticks. Game Box Controllers~\cite{buttons} are attachments for existing controllers to enable players who may simply need larger buttons or taller joysticks to play.  

All these systems and devices have their unique affordances and limitations. For example, Quadstick is the most popular video game controller for quadriplegics, though it will not work with the upcoming PS5~\cite{quad}. There are several games where it is not possible to map a physical option on the Quadstick to a game action because of the large number of game actions possible. Our software based design allows creating macros, commonly used by video game players, where a sequence of game actions (e.g.\ jump + turn left) can be mapped to a single facial expression. BCI interfaces require the user to wear a headset which may be difficult to wear and use for extended periods of time for playing games~\cite{vsumak2019empirical}. Our solution does not require the user to wear any sensors, trackers or devices. To our knowledge, facial expression recognition has not yet been investigated in the context of game interaction for quadriplegic individuals.

\begin{table*}[!t]
    \centering
    \begin{tabular}{|c|c|c|c|}
    \hline
        \multicolumn{3}{|c|}{\textbf{Facial AUs Combinations or Facial Expression}} &
        \textbf{Keyboard Key} \\
        \hline 
           AU6 $> 2.0$ $\bigotimes$ AU12 $> 2.0$ & Cheek Raiser + Lip Corner Puller&
            Happiness & 1\\
        \hline
        AU1 $\bigotimes$ AU4 $\bigotimes$ AU15 & \begin{tabular}{c}
             Inner Brow Raiser + Brow Lowerer \\
             and Lip Corner Depressor 
        \end{tabular}
       & Sadness & 2\\
        \hline
        AU9 $> 1.4$ $\bigotimes$ AU10 $> 2.0$ & Nose Wrinkler + Upper Lip Raiser & 
        Disgust & 3\\
        \hline
        AU2 $> 0.5$ $\bigotimes$ AU5 $> 1.5$ & Outer Brow Raiser + Upper Lid Raiser & 
        Wide Eyes & 4\\
        \hline
        AU7 $> 1.4$ $\bigotimes$ AU23 $> 1.0$ & Lid Tightener + Lip Tightener & 
        & 5\\
        \hline
        AU4 $\bigotimes$ AU25 $\bigotimes$ AU26 & Brow Lowerer +
        Lips Part + Jaw Drop &  & 6\\
        \hline
    \end{tabular}
    \vspace{1em}
    \caption{Facial AU combinations with their descriptions and their approximate equivalent facial expressions mapped to keyboard keys. The numbers in the first column are the thresholds for AU intensity values (no number means AU presence/absence binary value was used.) If each AU intensity in an expression combination is greater than the specified threshold value then the corresponding key press in that row is triggered.}
    \label{tab:mappings}
\end{table*}

\section{System Design}
Interaction design often focuses on creating solutions that are generalizable to a large portion of the population. In contrast, Assistive Technologies (AT) are usually designed for the individual. Prior research shows the best effects of an AT are seen when it is developed with and tested by the potential end users~\cite{vsumak2019empirical}. In this work we take the AT approach to create software game input system and three test games in collaboration with Atieh, a quadriplegic engineering graduate student in our lab. We follow a similar design approach working with Atieh as used by Lin et. al.~\cite{lin2014design} who created a game controller and a mouse for a quadriplegic teen. 

The fundamental idea behind our system design was to make the most out of the small muscle movements available to Atieh. In our first meeting, we learned that Atieh could voluntarily control only one finger so hand-based systems were not an option. Mouth-based systems like the Quadstick~\cite{quad} were not an alternative due to Atieh's limited jaw range of motion. Lastly, Atieh's prior experience with gaze-based systems was frustrating and thus that input modality was also discarded as an input method. After some more brainstorming done in a Zoom session with Atieh where we discussed different input methods and reviewed prior research, we settled on camera-based facial expression recognition as the input method. Using a camera-based system over other methods like Interferi~\cite{iravantchi2019interferi}] or Earfieldsensing~\cite{matthies2017earfieldsensing} which require the user to either wear a facemask or an ear-plug, makes our system usable with devices that most users are likely to already own, e.g., smartphone or a desktop webcam or a laptop with a built-in webcam.The next step was testing which facial expressions Atieh could make, the robustness of detecting those expressions, and the positioning of the webcam to capture their facial input in relation to the computer and display. The design and ideation for the system was conducted with Atieh over Zoom as social distancing rules made meeting in-person impossible.

\subsection{Pipeline} The system has four main parts: 1) facial expression recognition (FER) or facial action unit (AU) recognition, 2) speech recognition, 3) interaction design (AU and text to keyboard mapping), and 4) game design and gameplay. Figure~\ref{fig:pipeline} shows the system pipeline. Input from the webcam and the microphone is sent through the AU recognition system to the keyboard mapper, which converts them into keyboard inputs for each game. We used Unreal Engine (UE) version 4.25.1 for building the Walking Adventure (WA) and the First Person Shooter (FPS) games and UE version 4.23 for the Car Racing game. Game logic was created using UE’s Blueprint system.

% \begin{figure}[!b]
% \centering
% \begin{subfigure}[b]{.57\columnwidth}
% \centering
% \includegraphics[width=1\columnwidth]{figures/expressions_names.png}
% \end{subfigure}\hspace{0.25em}
% \begin{subfigure}[b]{0.38\columnwidth}
% \centering
% \includegraphics[width=1\textwidth]{figures/aloy-setup.png}
% \end{subfigure}
% \caption{Left: Six facial expressions used for playing the games. Top row left to right: Happy face, Sad face, Disgust. Bottom row left to right: Wide open eyes, Pucker, Jaw drop. Right: Atieh playtesting the FPS game at home.}
% \label{fig:usedExpressions}
% \end{figure}

\begin{figure}[!b]
  \centering
  \begin{tabular}{c@{}c}
  \begin{tabular}{@{}c@{}}
    \includegraphics[width=.56\columnwidth]{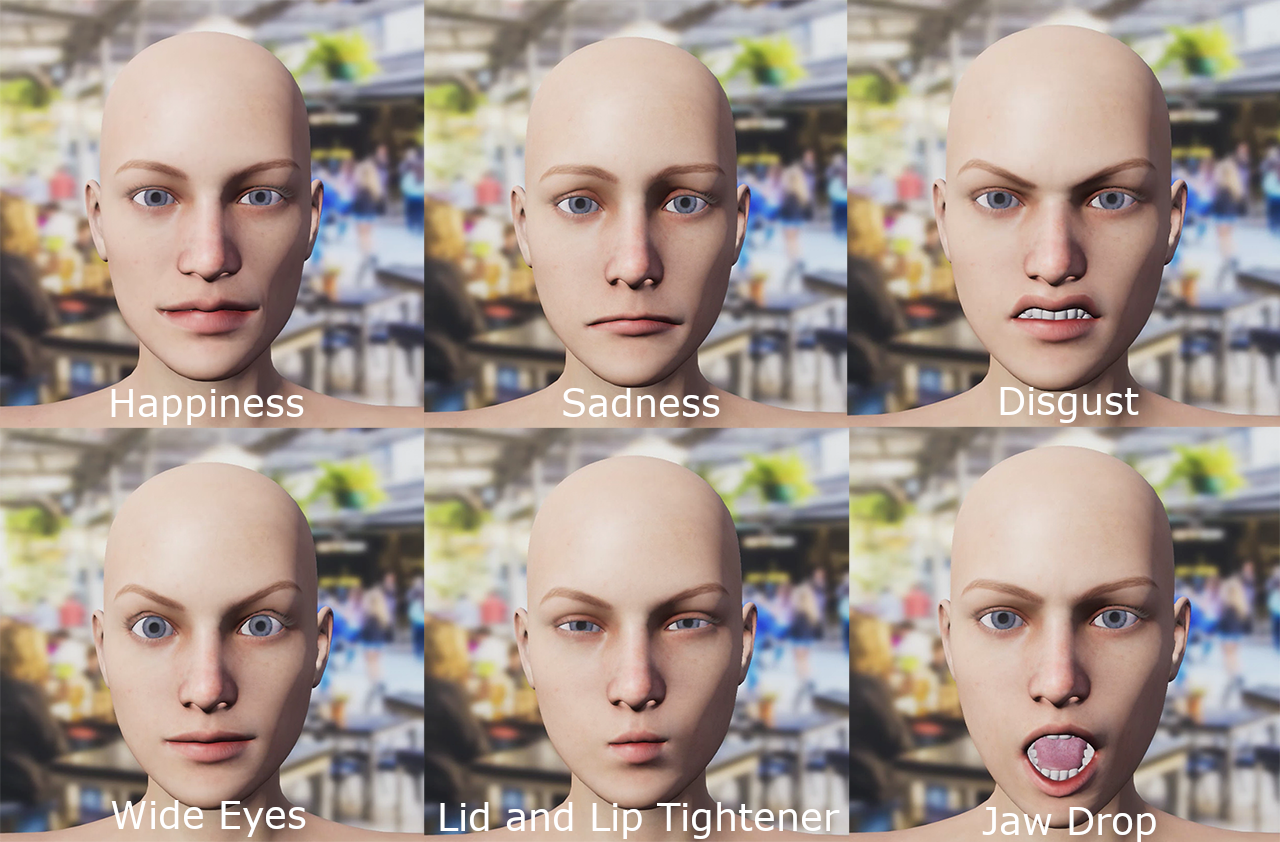} \\[\abovecaptionskip]
  \end{tabular}
    & \hspace{0.25em}
  \begin{tabular}{@{}c@{}}
    \includegraphics[width=.37\columnwidth]{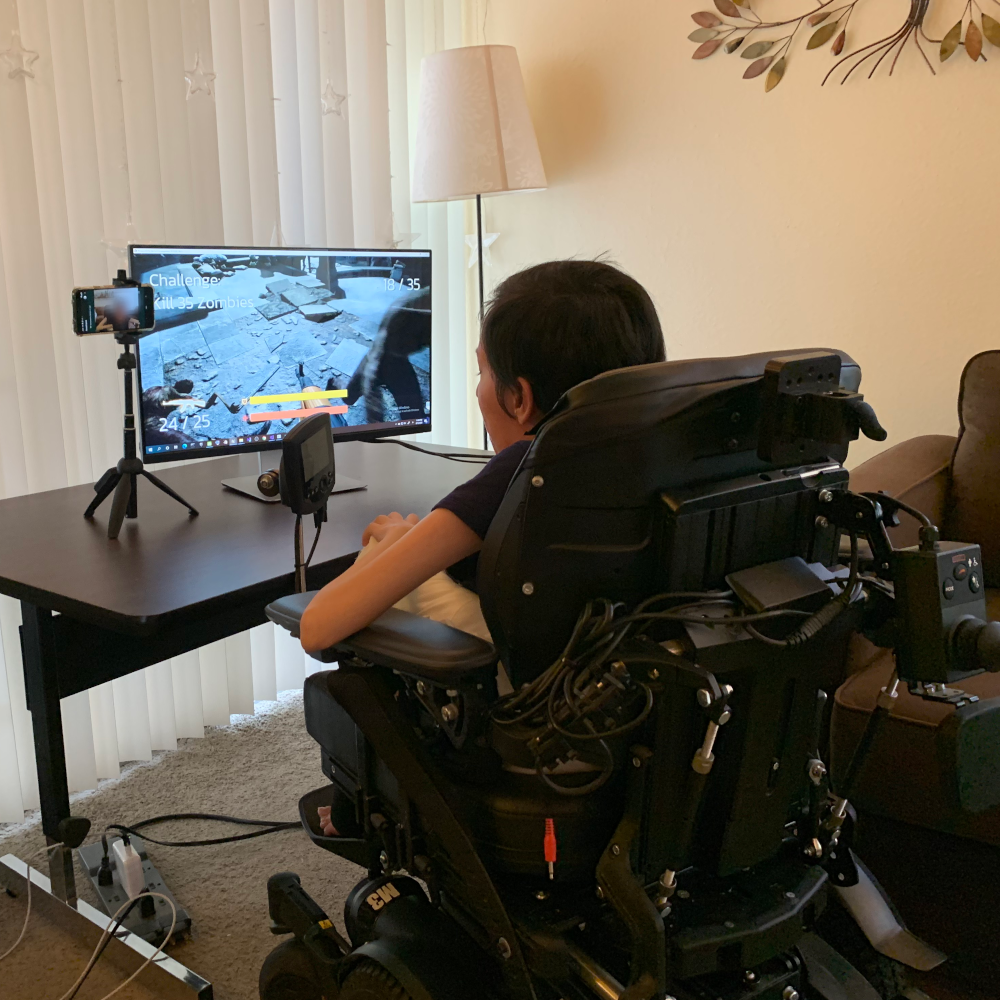} \\[\abovecaptionskip]
  \end{tabular}
  \end{tabular}
  \vspace{-1.5em}
  \caption{Left: Six facial expressions used for playing the games. Top row left to right: Happy face, Sad face, Disgust. Bottom row left to right: Wide open eyes, Pucker, Jaw drop. Right: Atieh playtesting the FPS game at home.}
 \label{fig:usedExpressions}
\end{figure}

 \begin{figure*}[!ht]
    \centering
    \includegraphics[width=.98\textwidth]{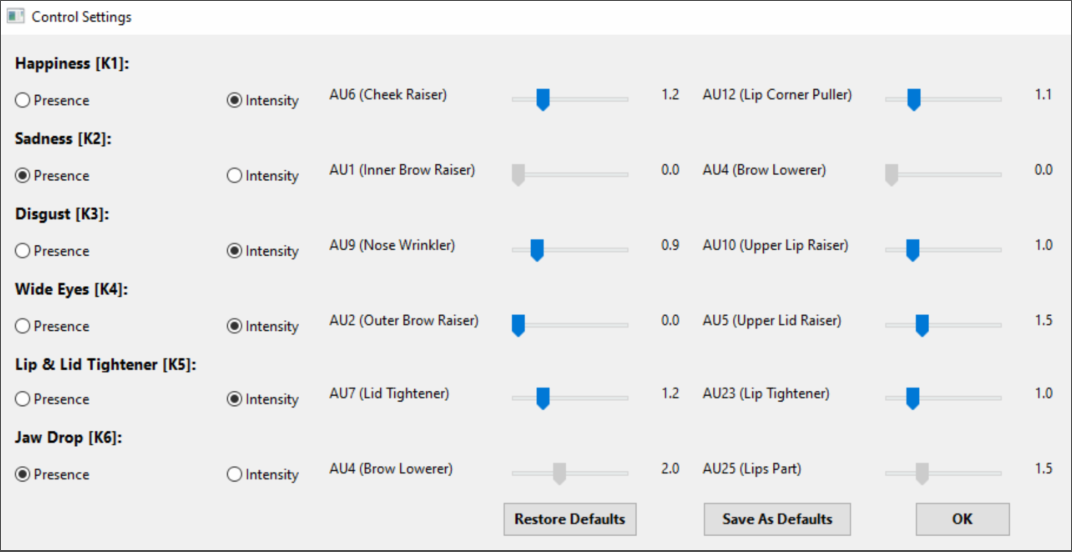}
    \caption{We built an interface to enable users to change AU detection thresholds and mappings for each game.}
    \label{fig:flex}
\end{figure*}

\subsection{Facial Input}\label{subsec:Sys-FER}
In order to standardize facial expression recognition, Ekman and Friesen~\cite{ekman1971constants} categorized facial muscle movements into \textit{Facial Action Units (FAUs)} to form the \textit{Facial Action Coding System (FACS)}. Over the years, two main types of approaches have been explored for FAU recognition - those that use texture information and those that use geometrical information~\cite{kotsia2008texture}. In our system, we use the OpenFace 2.0 toolkit~\cite{baltrusaitis2018openface} based on the texture approach, hereon referred to as OpenFace. The toolkit is capable of facial landmark detection~ \cite{zadeh2017convolutional}, head pose estimation, FAU recognition~\cite{baltruvsaitis2015cross}, and eye-gaze estimation~\cite{wood2015rendering}. Of these features, we integrated FAU recognition into our pipeline to extract Action Units (AUs) from the input video stream in realtime.

In our pipeline, AU detection takes two forms: 1) AU presence - a binary value depicting whether a particular AU is present in the captured frame or not, and 2) AU intensity - a real value between $0.0$ and $5.0$ depicting the intensity of the extracted AUs in a captured frame. OpenFace provides access to AUs 1, 2, 4, 5, 6, 7, 9, 10, 12, 14, 15, 17, 20, 23, 25, 26, 28, and 45. After testing, we discarded AUs 14, 17, 20 because they were similar to other AUs. AU45 detected blinking and was therefore unusable as input.

In each input frame, the FER system outputs AU presence and intensity values for each of the 18 AUs. These values are sent to the our keyboard mapper for converting them to game input. %(Table~\ref{tab:mappings}). 
Figure~\ref{fig:usedExpressions}: Left demonstrates the six FEs the player makes for taking actions in the games. Each one of these FEs is obtained by combining two or three facial AUs (Table~\ref{tab:mappings}). Figure~\ref{fig:usedExpressions}: Right shows Atieh playtesting the FPS game at home.

The AU combinations used in each game are experimentally determined by Atieh since it was not possible for them to make every facial expression that an able-bodied person might easily make. Additionally, some AUs had higher reliable detection than others. The experimentation involved testing with AU presence values first and replacing with AU intensity values if AU presence alone failed to be accurately detected. The numbers in Table \ref{tab:mappings} show the AU intensity threshold values for the AU combinations that were used in the games. These values are also experimentally determined with help from Atieh. All AU combinations use intensity values except two FEs; sadness, a combination of AU1 + AU4 + AU15, and jaw drop, a combination of AU4 + AU25 + AU26. These two FEs were reliably detectable with AU presence values alone.

\subsection{Speech Input}\label{subsec:Sys-SR}

We used a Python speech recognition library~\cite{SpeechRecognition01} to communicate with Google Cloud Speech API~\cite{GoogleSpeechRecognition} for converting spoken commands to text. The text data was scanned for specific keywords like ``Walk'' or ``Yes'' and converted into keyboard input using Pynput~\cite{pynput} and mapped to  keys previously programmed in Unreal Engine for each action in each game. Speech interaction served as a backup modality to AU recognition, always available for use but not necessarily required. 

\subsection{Key Mapping}
Facial expressions and text keywords are both mapped to keyboard input through the keyboard mapper. During testing, it became evident that AU recognition and mapping per input frame was frustrating to the user due to the system making multiple keyboard mappings per second. To resolve the issue, we set a threshold for the number of consecutive frames an AU combination needed to be visible in before getting mapped to the keyboard. This helped provide more control to the user and improved reliability.  After testing all AU combinations were set to a five frame threshold. 
Having a high frame threshold helped prevent the system from responding to unplanned facial muscle movements. 

\subsection{Flexibility}
While the mappings presented in this work are best suited for Atieh, we created an interface that allows the user to change the mappings as needed. For example, if a particular expression is not as robustly detected, or if the user finds the mapping of a smile to walking or running not intuitive for them, they can easily change which expression they would like to use through a front-end interface and map it to a game action of their choice (Figure~\ref{fig:flex}).

\section{Game Design and Gameplay}
There are more than 91 different video game genres~\cite{wikipedia}. We chose three different ones to implement and test our facial AU recognition based input system in games that are vastly different from one another. All games were collaboratively and iteratively designed with Atieh's feedback. For maintaining Atieh's privacy, we have removed their face from the gameplay in all the figures and replaced it with a virtual character wearing the same expressions. The mapping of AU combinations to keyboard keys is shown in Table ~\ref{tab:mappings}.

% \begin{figure}[!t]
% \centering
% \begin{subfigure}[b]{.46\columnwidth}
% \centering
% \includegraphics[width=1\columnwidth]{figures/3-walk-pickup.png}
% \end{subfigure}\hspace{0.25em}
% \begin{subfigure}[b]{0.46\columnwidth}
% \centering
% \includegraphics[width=1\textwidth]{figures/56-walk-turnleftright.png}
% \end{subfigure}
% \caption{Left: The player bending over to pick up a flower in Nature Walk with the requisite facial expression shown as inset. Right: Two different facial expressions or speech input allow the player to make the decision to turn left or not in Nature Walk.}
% \label{fig:walkpickup-walkturn}
% \end{figure}

\begin{figure}[!t]
  \centering
  \begin{tabular}{c@{}c}
  \begin{tabular}{@{}c@{}}
    \includegraphics[width=.46\columnwidth]{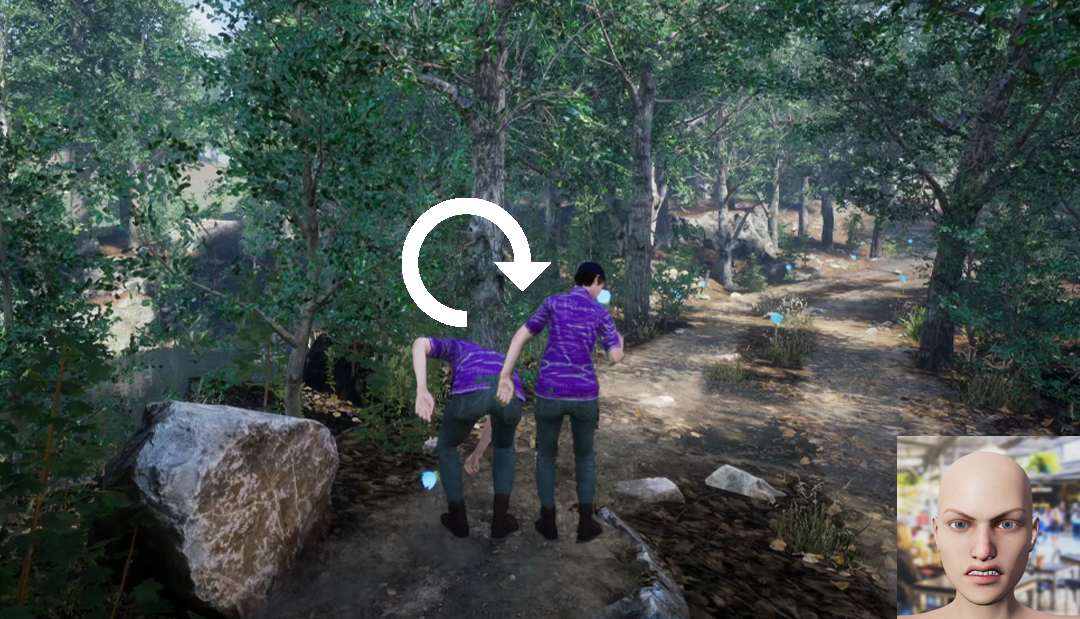} \\[\abovecaptionskip]
  \end{tabular}
    & \hspace{0.25em}
  \begin{tabular}{@{}c@{}}
    \includegraphics[width=.46\columnwidth]{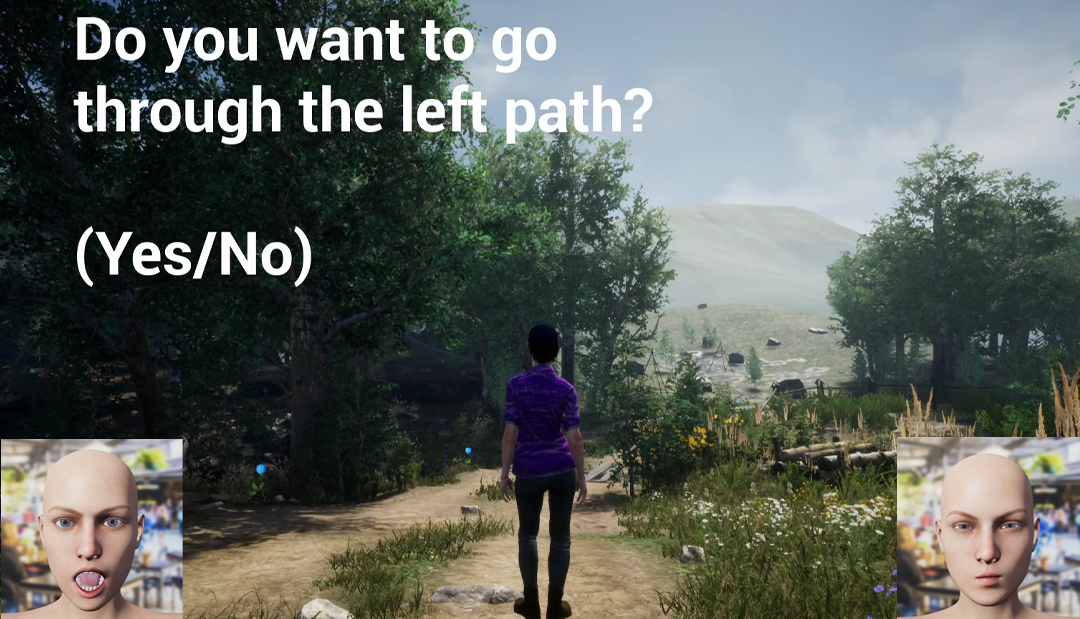} \\[\abovecaptionskip]
  \end{tabular}
  \end{tabular}
  \vspace{-1.5em}
  \caption{Left: The player bending over to pick up a flower in Nature Walk with the requisite facial expression shown as inset. Right: Two different facial expressions or speech input allow the player to make the decision to turn left or not in Nature Walk.}
 \label{fig:walkpickup-walkturn}
\end{figure}

\subsection{Walking Adventure}\label{sec:WA}
Knowing that Atieh has not played any video games due to being severely affected by SMA for years, we wanted to introduce them to the game controls and genres step by step, going from fewer to more interactions and from slower to faster paced games.  
Walking Adventure has three levels: Nature Walk, Cave Explorer and Temple Looter. The keyboard to game action mapping is presented in Table~\ref{tab:WalkingSim-table}: Left. Each level is visually different, designed to immerse the player in a different environment with its own set of tasks and goals. The game character is created in Adobe Fuse~\cite{AdobeFuse} and rigged and animated using Mixamo~\cite{Maximo}. All three games use ambient and task related sounds. Each time a facial expression is correctly detected, the system provides audio feedback to the player. Thus the player always feels in control of the game character's actions with multimodal feedback from the visual and the audio channels. 

\begin{table*}[!t]
%\begin{tabular}{|>{\centering}m{0.3\columnwidth}|>{\centering\arraybackslash}m{0.6\columnwidth}|}%{|c|c|c|}
\begin{tabular}{cc}
\begin{tabular}{|>{\centering\arraybackslash}m{.09\textwidth}|>{\centering\arraybackslash}m{.13\textwidth}|>{\centering\arraybackslash}m{.11\textwidth}|}
\hline
\textbf{Keyboard Key} & \textbf{Game Action} & \textbf{Input Type} \\ \hline
1                     & Start Walking        & AU\\ \hline
2                     & Stop Walking         & AU\\ \hline
3                     & Pick up              & AU\\ \hline
4                     & Sprint               & AU \\ \hline
5                     & Turn  Yes           & AU or Speech\\ \hline
6                     & Turn  No           & AU or Speech\\ \hline
\end{tabular} &
\begin{tabular}{|>{\centering\arraybackslash}m{.09\textwidth}|>{\centering\arraybackslash}m{.2\textwidth}|>{\centering\arraybackslash}m{.11\textwidth}|}
 \hline
\textbf{Keyboard Key} & \textbf{Game Action}     & \textbf{Input Type}  \\ \hline
1                     & Start/Stop Walking Fwd & AU\\ \hline
2                     & Aim and Shoot              & AU\\ \hline
3                     & Start/Stop Turning Left    & AU\\ \hline
4                     & Start/Stop Turning Right   & AU\\ \hline
5                     & Jump                       & AU\\ \hline
6                     & Pause                      & AU or Speech\\ \hline
\end{tabular}
\end{tabular}
\vspace{1em}
\caption{Left: Mappings of the keyboard keys to the actions defined in the Walking Adventure game. Right: Mappings of keyboard keys to the actions defined in the FPS game.}
\label{tab:WalkingSim-table}
\end{table*}

\subsubsection{Nature Walk}

Nature Walk is the first level in the Walking Adventure game. The goal is to enjoy walking and exploring the level and interacting with the environment. Nature Walk is a nature scene with realistic water, trees, and flowers that are organically scattered throughout the map built with assets from the Meadow Environment~\footnote{\url{https://www.unrealengine.com/marketplace/en-US/product/meadow-environment-set}}. The level design consists of tree lined walking trails with branching paths. The level uses a modular spline path for the character to follow at a fixed walking speed. The player can stop to interact with the flowers that use an emissive texture to make them stand out from the rest of the vegetation. Interaction with the flower (Figure~\ref{fig:walkpickup-walkturn}: Left) proceeds as follows:
\begin{enumerate}
    \item If the player is in the walking state, stop the player.
    \item Over a maximum of 2 seconds, turn the player to face the flower.
    \item Over another second move the player up to 90\% of the distance from its current position towards the rose, such that when the character bends over to pick up the flower, their hand intersects with the flower's stem giving it a more natural appearance.
    \item Play the pickup animation and partway through the animation sequence, delete the flower on the ground, and spawn an identical flower attached to the player's hand.
    \item Wait for the animation to end, then let the player go idle with the flower in the character's hand for 3-4 seconds.
    \item When the idle time ends, play the put-down animation.
    \item When the player is bent half-way, destroy the rose that is in the character’s hand to make it look like the rose was actually was put back down on the ground.
    \item Wait for the put-down animation sequence to end.
    \item Find a spot on the spline that is ~100 steps ahead of where the player was before they chose to interact with the flower.
    \item Make the character face and walk towards that spot.
    \item Finally, make the character face forward and continue walking along the trail.
\end{enumerate}

At path branches, we added a decision making option to enable the player to choose if they want to turn or continue walking along the main path (Figure~\ref{fig:walkpickup-walkturn}: Right). In keeping with the lighthearted mood of the level, there were no timers or tasks. 

% \begin{figure}[!t]
% \centering
% \begin{subfigure}[b]{.46\columnwidth}
% \centering
% \includegraphics[width=1\columnwidth]{figures/2-cave-stop.png}
% \end{subfigure}\hspace{0.25em}
% \begin{subfigure}[b]{0.46\columnwidth}
% \centering
% \includegraphics[width=1\textwidth]{figures/4-temple-sprint.png}
% \end{subfigure}
% \caption{Left: Player pauses to look around for crystals in Cave Explorer using the facial expression shown in the inset. Right: Player sprinting in Temple Looter with facial expression input shown in the inset.}
% \label{fig:cavestop-templerun}
% \end{figure}

\begin{figure}[!t]
  \centering
  \begin{tabular}{c@{}c}
  \begin{tabular}{@{}c@{}}
    \includegraphics[width=.46\columnwidth]{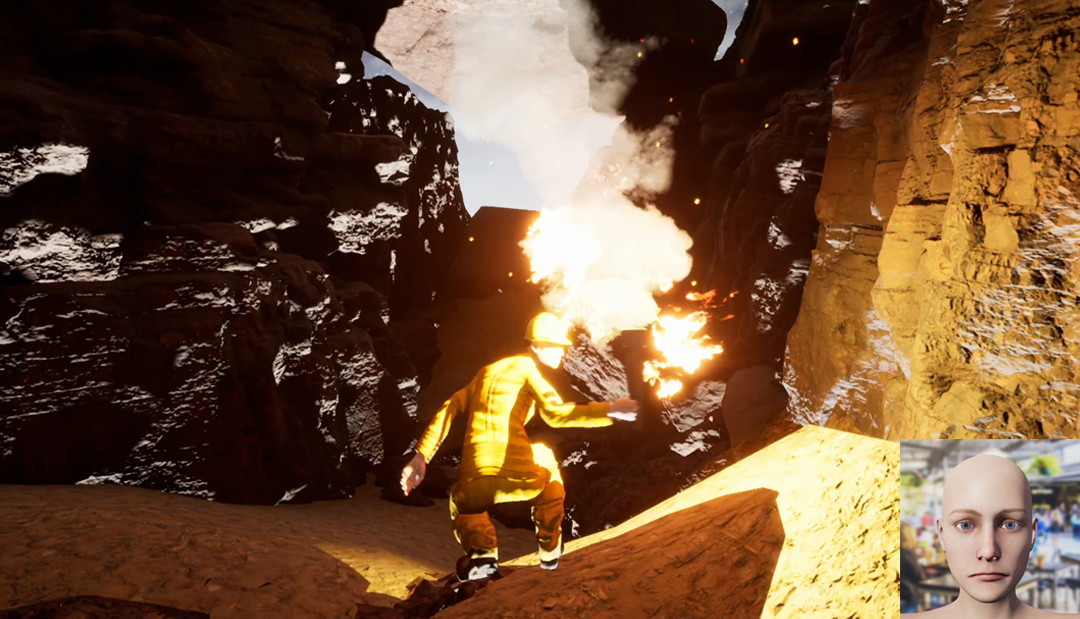} \\[\abovecaptionskip]
  \end{tabular}
    & \hspace{0.25em}
  \begin{tabular}{@{}c@{}}
    \includegraphics[width=.46\columnwidth]{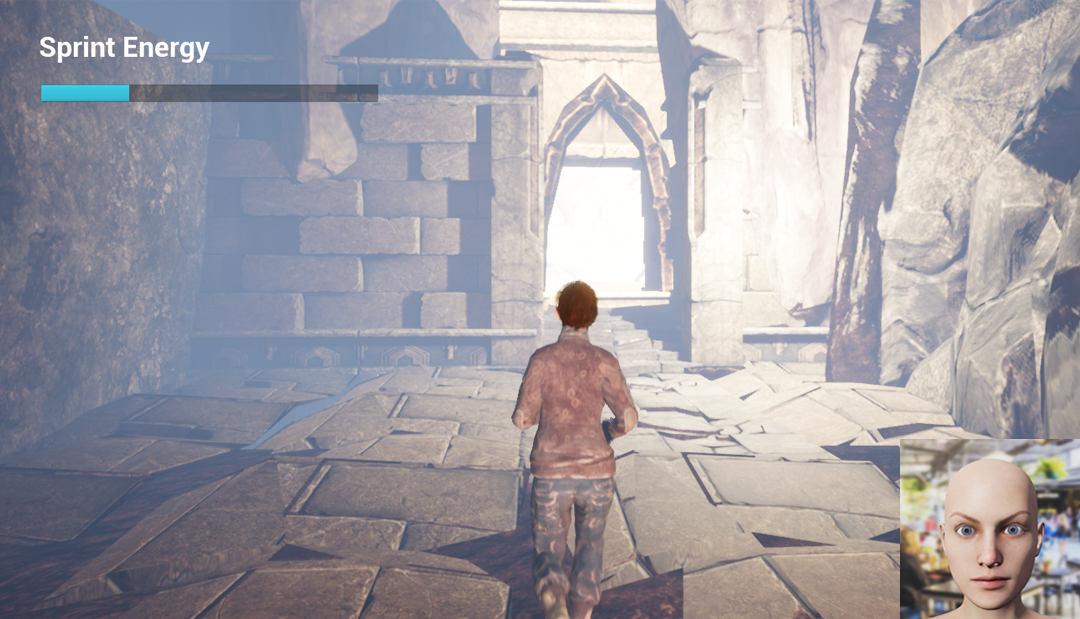} \\[\abovecaptionskip]
  \end{tabular}
  \end{tabular}
  \vspace{-1.5em}
  \caption{Left: Player pauses to look around for crystals in Cave Explorer using the facial expression shown in the inset. Right: Player sprinting in Temple Looter with facial expression input shown in the inset.}
 \label{fig:cavestop-templerun}
\end{figure}

\subsubsection{Cave Explorer}
For designing the Cave Explorer level, we downloaded free 3D assets from Quixel’s Limestone Quarry~\footnote{\url{https://quixel.com/megascans/collections?category=environment&category=natural&category=limestone-quarry}} collection. The downloaded assets were assembled to create a complex cave environment as shown in Figure~\ref{fig:cavestop-templerun}: Right The player explores a dimly lit cave bearing a torch in hand. Their task is to collect crystals and safely exit the cave. The crystal collection mechanic was added not only to mimic a task commonly found in commercial video games but also to make the level more interactive and goal oriented. The game character from Nature Walk is reused with different clothing and animations. To add an element of surprise, we implement a ``jump scare'' mechanic by hanging zombie skeletons from the cave roof that would abruptly drop down with appropriate animations and sound effects to scare the player at opportune moments.

\subsubsection{Temple Looter}
The last level in the Walking Adventure game is built by modifying a map in the free Infinity Blade: Fire Lands asset~\footnote{\url{https://www.unrealengine.com/marketplace/en-US/product/infinity-blade-fire-lands?lang=en-US}}. We created an ancient temple scene, similar to what one might see in an Indiana Jones movie. We again used the character from Nature Walk with an adventurer's clothing and different animations. The player is tasked with looting hidden treasure in the temple and escaping. We added a stamina bar that the player needed to fill up by not spending too much energy before sprinting out of the temple (Figure~\ref{fig:cavestop-templerun}: Left). Running is a new action added to this level, something that Atieh expressed a strong desire to be able to do.

Atieh's response following gameplay was,
\begin{quote}
    “The three levels of Walking Adventure have nice environments. I loved taking actions in the game world like jumping, picking up, walking around on it. I’d like the game better if there was a story behind, but I realize it’s only an exploratory design.”
\end{quote}

\subsection{First Person Shooter}

\begin{figure}[!b]
  \centering
  \begin{tabular}{c@{}c}
  \begin{tabular}{@{}c@{}}
    \includegraphics[width=.46\columnwidth]{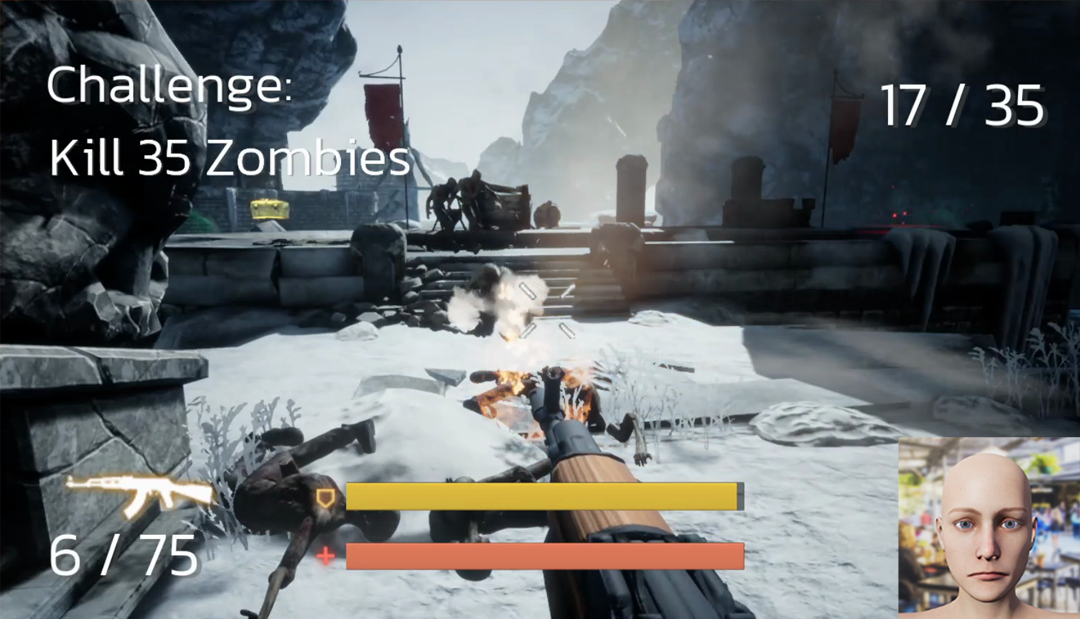} \\[\abovecaptionskip]
  \end{tabular}
    & \hspace{0.25em}
  \begin{tabular}{@{}c@{}}
    \includegraphics[width=.46\columnwidth]{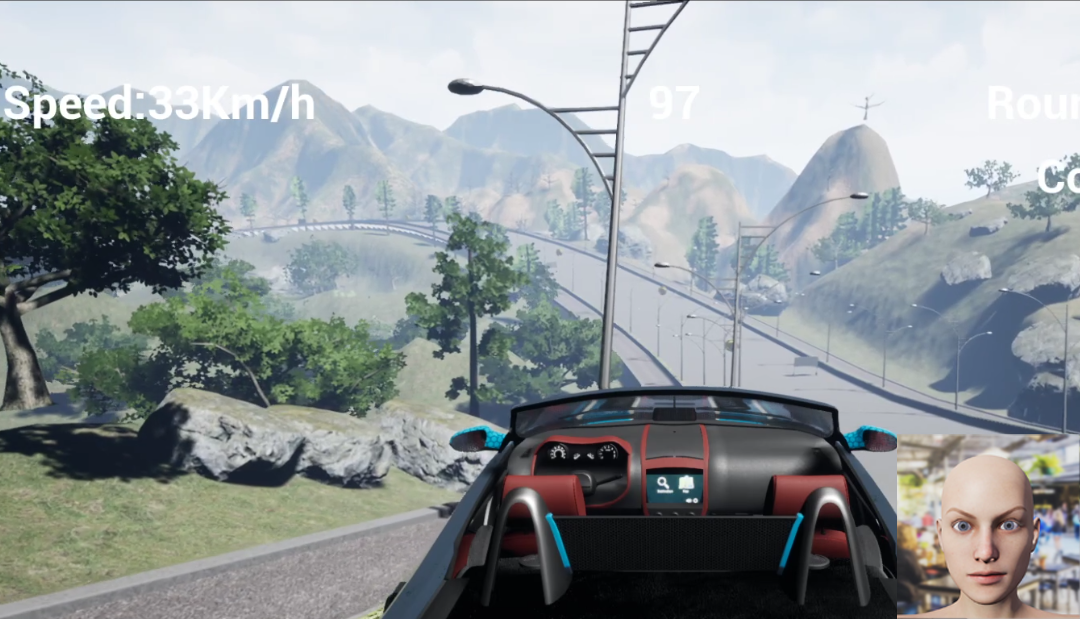} \\[\abovecaptionskip]
  \end{tabular}
  \end{tabular}
  \vspace{-1.5em}
  \caption{Left: Player shooting zombies in the FPS game using a facial expression shown as inset. Right: Player turning the car right using the facial expression shown as inset.}
 \label{fig:fpsshoot-carright}
\end{figure}

The second type of the game we created is a First Person Shooter (FPS). As opposed to Walking Adventure, which is a third-person game, the FPS allows Atieh to see through the eyes of the character. 
The FPS character was downloaded from Mixamo along with 40 animation sequences for covering typical movements in an FPS game. A blendspace was created to manage the animation logic playback with actions like walking, turning, jumping, aiming and shooting, reloading, and crouching. A single weapon option is added with sound effects and gunfire animation that showed as a flash at the tip of the gun (Figure~\ref{fig:fpsshoot-carright}: Left). The zombies from Cave Explorer are re-used with different animations to walk, attack, and die. Pathfinding logic is created for the zombies to move towards the player, if the player gets within a certain distance range. A horde system is implemented to spawn new zombies based on where the player is heading. We added trigger boxes on some paths in the environment to spawn a horde of zombies in a plausible location. This created the effect of there being more zombies than there actually were, which helped with performance and made the game feel higher action. 
The FPS map was created using elements from a free asset on the Unreal Engine Marketplace called Infinity Blade: Ice Lands\footnote{\url{https://www.unrealengine.com/marketplace/en-US/product/infinity-blade-ice-lands}}. Atieh's initial playtest revealed the game to be difficult for an unseasoned FPS player and too fast paced for using facial expressions or even speech as input. To help a first time gamer enjoy the FPS without frustration, the maximum number of zombies at any instant was limited to 15 and auto-aim was added. The auto-aim feature works as follows: when the player intends to shoot a zombie, the player character is turned by a calculated amount per frame and the scoped gun is pointed at the nearest zombie within a pre-specified range. Shooting takes place automatically after auto-aiming. The gun holds 25 bullets at a time and reloads automatically. Ammunition packs are strewn around the map. Like any FPS game, the player needs to manage the bullets to have enough in case a horde of zombies attack. Continuously holding down a key for character movement, as is common in traditional FPS games, was changed to using a key toggle. This meant the same facial expression could start and stop a character action like walking forward or turning left/right making it easier for the player to control the character. While speech input worked well for the slower paced Walking Adventure games, it was unusable for the FPS due to latency in cloud speech processing. Thus, speech was used only for pausing the game and not for the main actions. The facial expressions mapping to FPS actions is presented in Table~\ref{tab:WalkingSim-table}: Left.

About the FPS game, Atieh said, 
    
\begin{quote}
``[I]t is smooth and it did not frustrate me playing whereas most of the time any assistive tool that comes out for people with disability, would somehow need the person an exhausting effort. If you notice, for example when the character shoots the zombies, it is only a matter of lowering your lips and your eyebrows which is very simple. For me, it is a fun experience...I just wish it was a multi-level game.'' 
\end{quote}

% \begin{figure}[!b]
% \centering
% \begin{subfigure}[b]{.46\columnwidth}
% \centering
% \includegraphics[width=1\columnwidth]{figures/2-fps-shoot.png}
% \end{subfigure}\hspace{0.25em}
% \begin{subfigure}[b]{0.46\columnwidth}
% \centering
% \includegraphics[width=1\textwidth]{figures/4-car-right2.png}
% \end{subfigure}
% \caption{Left: Player shooting zombies in the FPS game using a facial expression shown as inset. Right: Player turning the car right using the facial expression shown as inset.}
% \label{fig:fpsshoot-carright}
% \end{figure}

\subsection{Car Racing}
The game is based on the Car Game template from Unreal to which 3D assets consisting of hills and trees were added from CSDN\footnote{\url{https://www.csdn.net/}}, a community of game developers. The player competes against a clock on a closed loop track.
To make the game more engaging, the player collects coins along the track, an idea inspired by the popular Nintendo car racing game, Mario Kart. The coins are created using simple cylinders with yellow material and spin animation attached. Other than coins, there are obstacles on the road that the player needs to maneuver around as bumping into them reduces the car's speed. During playtesting, Atieh found the default car speed to be too high to easily control with facial expressions. That made the car difficult to maneuver around  obstacles which required  many iterations of complex manipulations in a fast sequence. To help make the game more fun, the maximum car speed was lowered, the obstacle physics was simplified to enable turning the car easily with a facial expression toggle, similar to turning in the FPS game (Figure~\ref{fig:fpsshoot-carright}: Right). The mapping of facial expressions to car motions is presented in Table~\ref{tab:RacingCar-table}.

After playing the Car Racing game, Atieh said,
\begin{quote}
    ``In the past, when I could still play with my hands and regular controllers, I loved playing racing games. However, losing my ability to use my hands, playing this genre of game became impossible. I could never have imagined that I would be able to play my favorite game genre again without putting in too much effort. When the idea of incorporating facial expressions in this genre first came up, I thought well, if we could make the game so customized, we could probably make it possible. Though surprisingly, after only the second attempt that lowered the vehicle speed, the game resembled what I used to play years ago and became very entertaining.''
\end{quote}

\begin{table}[!t]

\begin{tabular}{|c|c|c|}
\hline
\textbf{Keyboard Key} & \textbf{Game Action}        & \textbf{Input Type}\\ \hline
1                     & Start/Stop Driving Forward  & AU\\ \hline
2                     & Start/Stop Driving Backward & AU\\ \hline
3                     & Start/Stop Turning Left     & AU\\ \hline
4                     & Start/Stop Turning Right    &AU \\ \hline
\end{tabular}
\vspace{1em}
\caption{Mappings of the keyboard keys to the actions defined in the Car Racing game.}
\label{tab:RacingCar-table}
\end{table}

\section{Design Reflection}

Designing our exploratory system was achieved iteratively through conducting a series of experiments with Atieh. Each stage of the design process gave us new insights into changing configuration of the expressions and game actions in ways that would best match the player's abilities. While our current design is focused on Atieh specifically, the system is easily modifiable. In fact, Atieh was able to create macros for testing the system with commercial games (Section~\ref{sec:limbo}), which is something we had not used with the games we created. Through our process of weekly interactions with Atieh over 6 months, we learned some valuable lessons that may help others design FER based input systems for quadriplegic users. 

\subsubsection{Design Input to Support Player Ability} While obvious, determining player ability is the first step to figuring out a potential input method that would work best for that individual. Our interaction system primarily focuses on using facial muscle movements because that's what we narrowed down options to during our first conversation with Atieh. Subsequently, we learned that the number of muscles employed in each facial expression also makes a big difference to the player's comfort level and to the system's ability to detect expressions reliably. Using too many muscles repetitively is tiring and difficult to control while using too few leads to false positives. Atieh tried several facial expressions to help us establish that expressions using 2-3 AUs worked best for them for two main reasons: 1) easy to make repeatedly, and 2) reliably detected without false positives.

\subsubsection{Consider Frequency of Game Actions}
There are some actions in every game that are more frequently taken than others. In our games they were: start/stop walking in the Walking Adventure, aiming and shooting in the FPS, and turning left or right in the Car Racing game. Keeping their frequent usage in mind, the facial expressions selected for these actions needed to be less demanding on the player. While the mappings used in the current games are comfortable for Atieh, they are also easily changeable if Atieh's abilities change over time (Figure~\ref{fig:flex}). For Atieh, the four FEs (i.e.\ happiness, sadness, disgust, and wide eyes) were easier to make compared with the others and of these four, disgust and wide eyes were the simplest. Therefore, we mapped \textbf{disgust} and \textbf{wide eyes} to the most frequent game actions. We simultaneously attempted to map positive expressions with positive game actions, for example, smiling to move forward, to help the user remember the mapping.

\subsubsection{Consider Multimodal Input and Feedback}
Video games come in a large variety of genres. Some require fast responses while others are more forgiving. Relying on a single input system can limit the types of interactions and games that may work successfully. Including multiple input modalities, like FER and speech in our system, can help provide the user a backup input system for situations when the primary system is difficult to use. The inverse is also possible in that some games may not work well with one modality but work flawlessly with another. An example of this is our FPS game that would not work well with speech input due to latency induced by cloud-based recognition while the Walking Adventure game was perfectly suited for speech input. Similarly, including multimodal feedback through visuals, text and sound data can help the player stay in control, knowing that their expression was recognized. 

\subsubsection{Design Games for Personalization and Flexibility}
Allowing the games to be personalized can help users enjoy playing them without getting frustrated. For example, mapping a specific expression to a game action,  slowing down or speeding up a particular game action, reducing or increasing the frequency of enemies are elements that the user should be able to configure at the start. Commercial games provide this feature in the form of Easy, Normal or Difficult modes of gameplay and having this choice is helpful for players of all abilities. 
Unfortunately, a large percentage of ATs for quadriplegic individuals are left unused for reasons from device performance to changes in user preference~\cite{schalk2008two, hurst2013making}. Designing with a quadriplegic user helped us better understand the constraints and opportunities. Our system is flexible and configurable (Figure~\ref{fig:flex}) such that the player can map any expression to any keyboard key and also create macros to allow for multiple key presses in sequence for more complex game input. 

After months of playtesting and building, when Atieh played through the full games:
    
\begin{quote}
``All these years that I have had lost my ability to use my hands to grab game controllers or even since when I realized that using a keyboard and a mouse have become impossible I desperately looking for finding a way to regain those things that one day were my only way of having some entertainment in my life and the more I pursued the more I became disappointed and eventually gave up. Even those accessible tools that came out like eye-gazing devices could not give me back the ability to play the video games, and it gradually became like a dream for me to play and I started only watching other people's gameplays. But trying out this tool gave me hope that I still can go into the game world once again.`` 
\end{quote}

\subsection{Adoption in Other Video Games}\label{sec:limbo}
We were curious to test our system with some commercial games to explore how it would perform outside our designed games. We were surprised to find that games that do not require a mouse or games where the mouse movements can be replaced with keyboard input, worked well with our system without requiring any modifications. We tested four atmospheric single player puzzles games on Steam like GRIS\footnote{\url{https://store.steampowered.com/app/683320/GRIS/}} and Inside\footnote{\url{https://store.steampowered.com/app/304430/INSIDE/}}, Limbo\footnote{\url{https://store.steampowered.com/app/48000/LIMBO/}}, and Little Nightmares\footnote{\url{https://store.steampowered.com/app/424840/Little_Nightmares/}}. For all these games, we created macros for game actions that required rapid key presses in a particular sequence to accomplish the game task. With that small change, the games were fully playable without requiring Atieh to make multiple facial expressions in quick succession. To switch to the macro mode and back, a FE was used. Based on this experiment, Atieh is excited about the potential of our system to work with other commercial games and they will continue testing. We believe that the set of games that work with our system will grow as more developers enable mapping game controls to a keyboard.    

\subsection{Limitations and Future Work}
Our work contributes to a body of research and design of hands-free gaming input for users with severe motor disabilities who need innovative solutions that can enable them to play independently. Our work identifies potential for future research in the field of hands-free input based on facial expression recognition. 
Because the current system was designed in collaboration with one quadriplegic individual, we did not build an interface to enable mapping expressions to keys. A future version will be helped with such an interface allowing players to customize the system according to their needs. An important future direction is improving the user interface, allowing for personalization and flexibility. Additionally,  automated adaptation to disease progression would enable the system to continue being meaningful to the user for a long time. Once campus COVID-19 restrictions\footnote{\url{https://www.cdc.gov/coronavirus/2019-ncov/community/colleges-universities/index.html}} are lifted, conducting a user study with a group of quadriplegic individuals would help us understand and modify the system design to improve flexibility and usability for individuals who are different from Atieh. The study would include both the quantitative methods, as well as the qualitative methods, to understand challenges in adopting a new AT and the value it brings.

\section{Conclusion}
In this paper, we explored the design of a hands-free interaction system for playing video games in collaboration with a quadriplegic player. We demonstrated the system in use with three different types of games. We found that the system enable Atieh to play and enjoy the experience without any frustration. This was the first time Atieh was able to play because they are unable to use input devices like the Quadstick~\cite{quad}. Since every motor impaired individual has unique requirements, we believe our software solution can be easily customized for their abilities and purposes, with the assumption that the individual has voluntary control over their facial muscles. With a growing number of game developers and companies including accessibility options into the games, we are hopeful that facial expression recognition based input would also become an option soon, opening up a new world of gaming for Atieh and perhaps many others. 

\begin{acks}

% \section{Acknowledgments}
We would like to thank Shiran Wang for work on the Car Racing game. We would like to acknowledge members of the Perceptual Engineering Lab in the Computer Science department at UCSB for their helpful comments on our work. 

\end{acks}

\bibliographystyle{ACM-Reference-Format}
\bibliography{sample-base}

%\appendix

\end{document}